\documentclass[]{emulateapj}

\usepackage{natbib}
\usepackage{graphicx, amsmath, amsthm, amssymb}

\newcommand{\Msol}{M_\odot}

\newcommand{\MJ}{M_{\rm J}}

\newcommand{\MB}{M_{\rm B}}

\newcommand{\Mearth}{M_\oplus}

\newcommand{\aB}{a_{\rm B}}
\newcommand{\mB}{m_{\rm B}}
\newcommand{\eB}{e_{\rm B}}
\newcommand{\IB}{I_{\rm B}}
\newcommand{\PB}{P_{\rm B}}
\newcommand{\Ikoz}{I_{\rm koz}}

\newcommand{\Ham}{{\cal F}}
\newcommand{\Hamp}{{\cal F}^\prime}
\newcommand{\Hamq}{{\cal F}_{\rm q}}

\newcommand{\tppin}{\tau_{\rm pp,1}}
\newcommand{\tppout}{\tau_{\rm pp,2}}
\newcommand{\tkozin}{\tau_{\rm koz,1}}
\newcommand{\tkozout}{\tau_{\rm koz,2}}

\newcommand{\tppnode}{\tau_{\rm \Omega pp}}

\newcommand{\Lin}{{\bf L_1}}
\newcommand{\Lout}{{\bf L_2}}
\newcommand{\LB}{{\bf L_{\rm B}}}

\newcommand{\be}{\begin{equation}}
\newcommand{\ee}{\end{equation}}
\newcommand{\bea}{\begin{eqnarray}}
\newcommand{\eea}{\end{eqnarray}}

\shorttitle{Planetary Systems in Binaries. I.}

\slugcomment{to appear \apj}

\shortauthors{Takeda, Kita \& Rasio}

\begin{document}

\title{Planetary Systems in Binaries. I.  Dynamical Classification}

\author{Genya Takeda\altaffilmark{1},
Ryosuke Kita\altaffilmark{1},
and Frederic A. Rasio\altaffilmark{1}
}
\altaffiltext{1}{Department of Physics and Astronomy,
Northwestern University, 2145 Sheridan Road, Evanston, IL 60208; {\tt g-takeda, r-kita, rasio@northwestern.edu}}

\begin{abstract}
Many recent observational studies have concluded that planetary systems  commonly exist in multiple-star
systems.  At least $\sim20\,\%$ of the known extrasolar planetary systems are associated with one or more
stellar companions.   The orbits of stellar binaries hosting planetary systems are typically 
wider than 100\,AU and often highly  inclined with respect to the planetary orbits.
The effect of  secular perturbations from such an inclined binary orbit on a coupled system of 
 planets, however, is little understood theoretically.
In this paper we investigate various dynamical classes of double-planet systems in binaries 
through numerical integrations and we provide an analytic framework based on secular perturbation theories.    
Differential nodal precession of the planets is the key property that separates 
two distinct dynamical classes of multiple planets in binaries: (1) dynamically-rigid systems 
in which the orbital planes of planets precess in concert as if they were embedded in a rigid disk, 
and (2) weakly-coupled systems in which the mutual inclination angle between initially coplanar planets grows 
to large values on secular timescales.  In the latter case, the quadrupole perturbation from the 
outer planet  induces additional Kozai cycles and causes the orbital eccentricity of the 
inner planet to oscillate with large amplitudes.  The cyclic angular momentum transfer from 
a stellar companion propagating inward through planets can significantly alter the orbital 
properties of the inner planet on  shorter timescales.  This perturbation propagation
mechanism may offer important constraints on the presence of additional planets in known single-planet
systems in binaries.

\end{abstract}

\keywords{celestial mechanics -- binaries: general -- planetary systems}

\section{Introduction}

  As of February 2008, among the 220 planet-hosting stars (excluding pulsars) 44 (20\,\%) of them
are members of binary or higher-order multiple-star systems \citep{raghavan06,desidera07,eggenberger07}. 
Since the systematic adaptive-optics searches for stellar companions around known planetary systems 
are not yet complete, the true multiplicity among planet-hosting stars should be greater than 
20\% \citep{eggenberger07}.  Many of the stellar companions around planetary systems are
faint, distant, or both.  Nearly a dozen of them are K-type or later (including eight M dwarf companions 
and one brown dwarf companion).   Their  sky-projected  separations range from 20 to 12000\,AU
from the planetary systems, with a median of $\sim510\,$AU.  
This occurrence of planets in relatively wide binaries is certainly affected by 
the observational biases.   Spectroscopic binaries with angular separations 2--6$\arcsec$ 
are often deliberately excluded from the planet-search target samples since it is difficult
to extract precise radial-velocity information from the recorded flux contributed from the 
two components \citep{valenti05,eggenberger07}.  On the other hand, it is also shown by theoretical planet
formation simulations that nearby stellar companions may hinder planet formation processes 
in multiple ways.  First, a binary companion at a distance $\sim 20$\,--\,100\,AU tends to tidally 
truncate the nascent circumstellar disk around the primary, thereby reducing
the mass and lifetime of the disk \citep{artymowicz94,pichardo05,kley07}.   Periodic tidal heating
in the disk caused by a companion would also lead to the decay of spiral structures and 
evaporation of the volatile materials \citep{nelson00,mayer05}.  Even if planetesimals successfully form
in the disk, a close stellar companion would stir up their relative velocities so that they are likely 
to fragment through collisions rather than to coagulate \citep{kley07,paardekooper08,thebault06,marzari00}.  
From most of these studies it is 
inferred that the orbital separations of stellar binaries containing planetary systems can be preferentially 
larger than the binary systems without planets.

\begin{figure}
  \begin{center}
    \plotone{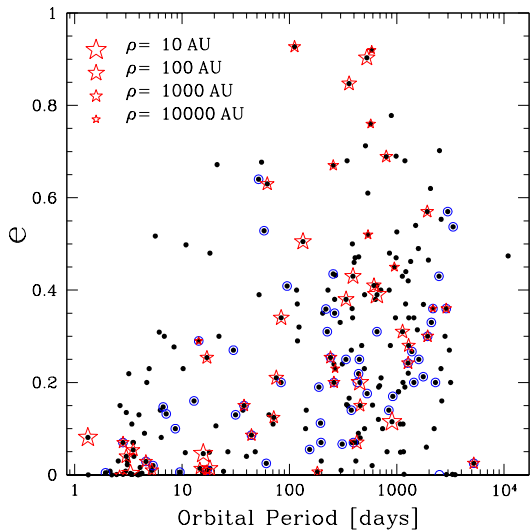}
    \caption{Eccentricity--period diagram for 234 extrasolar planets (black dots).  The red-star symbols 
	represent planets that are associated with binary companions, with the symbol sizes  scaled to
	$1 / \log{\rho}$, where $\rho$ is the projected binary separation.
	The circled-dots are the members of multiple-planet systems.
	\label{e_P.fig}}
  \end{center}
\end{figure}

Distant (and often faint) stellar companions, however, may still significantly alter the planetary orbits
around the primary stars on secular timescales.  \citet{zucker02} were the first to investigate
the statistical differences in the orbital properties of planets in binaries possibly caused by 
such secular perturbations.  The most distinctive trend they found was the occurrence of massive 
close-in planets exclusively among binaries.  Though they had only 9 planets in binaries in their samples
at that time, this trend has been robust for planets with $m\sin{i} > 2\MJ$ and $P < 100\,$days 
as the sample size increased  
\citep[see ][ and references therein]{udry07}.  There are also discernible differences in 
the eccentricity--period diagram for different populations of extrasolar planets  (Figure~\ref{e_P.fig}).
It has been pointed out by \citet{eggenberger04}  that planets in multiple-star systems show 
systematically lower orbital eccentricities if their orbital periods are less than 40\,days.  
On the other hand, there is a relative paucity of planets in binaries on near-circular orbits 
($e < 0.1$), for planets with orbital periods greater than 40\,days.
Indeed, in this region  the median orbital eccentricity of planets in binaries 
is 0.36, much higher than the median eccentricity 0.27 of planets around single stars.  
And lastly, the four largest orbital eccentricities and $\sim50\,\%$ of the 18 planets 
with $e>0.6$ are associated with confirmed binary companions.   These statistical trends
strongly suggest that the orbital evolution of planetary systems is commonly affected
by distant stellar companions.  

Theoretical studies have investigated the role of secular perturbations by stellar companions in 
the dynamical evolution of planetary systems using  simplified models.   
The Kozai mechanism, originally discovered by \citet{kozai62} for the orbital analysis of 
a Jovian asteroid under the perturbation from the Sun, has been adopted
for extrasolar planets and successfully modeled some of the highly eccentric planetary orbits, 
assuming that those systems contain only one planet \citep{holman97,wu03}.  
The Kozai mechanism permits the planet's orbital eccentricity to stably grow to arbitrarily 
large values up to unity, by tuning the unknown mutual inclination between the planetary and
binary orbits.  Recently, there is growing attention on the combined effect of the Kozai mechanism
and tidal dissipation mechanism as an alternative means to produce hot Jupiters  
\citep{wu03,fabrycky07,wu07}.  For the cases of multiple-planet systems in binaries,  
\citet{marzari05a}  have systematically studied the dynamical outcomes of initially-packed
triple-planet systems in binaries, for  limited cases in which the planetary systems are 
coplanar with the binary.

The secular effect of a non-coplanar stellar companion to a coupled system of multiple planets  is, 
however, an important problem which has  received  little attention in the past. 
In reality, planetary systems emerging from the nascent circumstellar disks are more likely to contain 
multiple planets \citep{ida05,thommes07}.  It is also rather unlikely that newly formed  multiple
planets orbit in the  plane of the binary.   \citet{hale94} has systematically studied 
binary systems with solar-type components in search for the alignment of their spin axes,
by combining the $v \sin{i}$ measurements with rotational period information.  
His results showed that spin alignments exist only for relatively close binary systems with 
orbital separations 30--40\,AU or less.  Because the binary separations observed among 
planet-hosting stars are typically much beyond this limit, 
planetary systems formed along the primary's equator are randomly inclined with respect to the 
binary plane.  In other words, the distribution of the mutual inclination angle $I$ between
the planetary orbits and the binary plane is uniform in $\cos{I}$. 
This implies that for about three-quarter of the cases planetary orbits are inclined with respect to the
binary plane by more than $40^\circ$, the critical Kozai angle ($\S\,$\ref{kozai}).

The effects of  Kozai-type perturbations to multiple-planet systems were first investigated 
by \citet{innanen97}.  While numerically studying the evolution of the Solar System
under the influence of a hypothetical stellar companion orbiting at 400\,AU from the Sun, 
they  discovered that, under certain circumstances, the orbital elements of the giant planets evolve in concert, 
as if they were embedded in a rigid disk.   \citet{malmberg07} have performed similar 
numerical experiments but with slightly different binary parameters.  In their simulations,
the orbital evolutions of the planets were decoupled, and the excited orbital eccentricities of 
the planets  induced instability and led to ejection of one or more  planets. 

The rigid coupling between planets in response to the binary perturbation found by \citet{innanen97}
has not been understood analytically.  A significant challenge is involved in developing a
generalized perturbation theory for $N$--body systems with $N > 3$  
since the approximation techniques employed in the traditional secular theories are no longer valid.
Higher order terms in the orbital eccentricities, inclinations, and semi-major axes 
may play important roles in the long-term evolution of multiple planets in binaries  
and thus cannot be eliminated from the disturbing function as  in typical secular theories. 
Systematic mapping of dynamical stabilities using numerical integrations is also 
impractical due to the large number of dimensions of the parameter space (the mass and the three-dimensional 
orbits for each planet and the binary).

The objective of this paper is to investigate the secular evolution of two planets in wide binaries
with a more generalized set of initial conditions and to provide analytic criteria that separate
 possible dynamical outcomes.   To disentangle the complex interplay between binary perturbations
and gravitational coupling between planets, we first separate the problem into 
two sets of secular three-body problems: the secular evolution of hierarchical, highly inclined
triple systems (the Kozai mechanism, \S~\ref{kozai}) and the secular evolution of  
isolated, near-coplanar double-planet systems (the Laplace-Lagrange secular theory, \S~\ref{LLTheory}).
In \S~\ref{dynamicaloutcomes}, we list the  important dynamical classes of double-planet
systems in binaries along with numerical simulations.  
The analytic explanation for each of the dynamical classes is presented in \S~\ref{discussion}.


\section{Secular Three-body Problems}
In this paper we focus on hierarchical four-body systems in which two  planets orbit a component of 
a wide stellar binary system.  The binary separation is typically on the order of $100\,$AU.  
The initial planetary orbits are assumed to be nearly coplanar and arbitrarily inclined with respect to the 
binary plane.  

To analyze the long-term evolution of hierarchical four-body systems, we first consider the gravitational
perturbation from the stellar companion and the mutual perturbations between the planets
separately.  The former has been studied in detail in the context of the Kozai mechanism, 
which takes account of full three dimensional orbits, assuming the hierarchy of the system.  
The latter is the secular evolution of a pair of planets around a single star, analytically
formulated through the Laplace-Lagrange theory, which does not assume any hierarchy but 
uses the assumption of small orbital inclinations and eccentricities.  
Both theories are developed by collecting only the secularly varying terms from the time-averaged
disturbing function.    This is an important point, indicating that these secular perturbations are generally
not cumulative; a secular perturbation mechanism occurring on a longer timescale
is averaged out to zero by the other perturbation (there are rare cases of secular resonances 
between two different  perturbation mechanisms, discussed in \S~\ref{rigidrotation}).  
In the following sections we summarize the relevant properties of Kozai-type perturbations and 
the Laplace-Lagrange secular theory, and derive the secular perturbation timescales of each mechanism.

Throughout this paper, subscripts 0, 1, 2, and B are used to refer to the properties of the 
primary star, the inner planet, the outer planet, and the binary companion, respectively.

\subsection{The Kozai Mechanism \label{kozai}}

The secular perturbation theory for a hierarchical, non-coplanar triple system was
first developed by \citet{kozai62}  in the analysis of the motion of a 
Jovian asteroid under the gravitational attraction of the Sun.  
Here we adopt the Kozai mechanism for the orbital evolution of a planet around a component of 
a wide binary system.  

The complete Hamiltonian of the three-body system is 
\begin{equation}
\Ham = - \frac{G m_0 m_1}{2 a_1} - \frac{G (m_0 + m_1) \mB}{2 \aB} + \Hamp. \label{Ham}
\end{equation}

The formulation of the Kozai mechanism takes advantage of the hierarchy of 
the system such that $\alpha \equiv a_1/\aB\ll 1$ and expands the perturbing Hamiltonian $\Hamp$ 
in terms of  $\alpha$ (since $\alpha$ is small, $\Hamp$ is a rapidly converging series).  
The  first term of  $\Hamp$ is the quadrupole-order potential $\Hamq$ \citep{ford00}. 
To derive the secular equations of motion, short-period terms are eliminated through averaging 
the Hamiltonian over the mean anomalies, $l_1$ and $l_{\rm B}$.
The averaged quadrupolar Hamiltonian is \citep{fabrycky07},
\bea
\bar{\Hamq} & = & -\frac{G m_0 m_1 \mB}{m_0 + m_1} \frac{a_1^2}{8 \aB^3 (1-\eB^2)^{3/2}}  \\
            &   &  \times \left[2+3e_1^2-(3+12e_1^2- 15e_1^2 \cos^2{\omega_1}) \sin^2{I}\right].  \label{ave_Hamq}
\eea
Since the binary orbit carries the majority of the angular momentum of the system, it is typically taken 
 as the reference plane ($I_{\rm B}=0^\circ$) that is invariant in the quadrupole approximation.  
Thus the angle $I$ in Equation~\ref{ave_Hamq} corresponds to the relative inclination of 
the planetary orbit measured from the reference plane.
With the averaged Hamiltonian, the Lagrange equations of motion can be derived (Equations~5 of 
Innanen et al. 1997) which can be numerically integrated for $e_1(t), I_1(t), \omega_1(t),$ and 
$\Omega_1(t)$.

Now we list the key properties of the Kozai Mechanism.

(1) Since the mean anomaly $l_1$ is deliberately removed from the averaged
Hamiltonian, its canonical conjugate momentum $L_1 \equiv m_0 m_1 \sqrt{G a_1 / (m_0+m_1)}$
and thus also the semi-major axis are constants of motion.   Because the averaged Hamiltonian
is independent of $\Omega_1$, its canonical conjugate $H_1$, corresponding to  the z-component angular
momentum, is also constant.  This implies that the averaged Hamiltonian is axisymmetric.  
Another integral of motion,  called the Kozai integral, can be  defined as 
\begin{equation}
\frac{H}{L} = \sqrt{1 - e_1^2}\cos{I} \equiv \sqrt{h} \label{kozaiintegral}.
\end{equation}
The Kozai integral couples the evolution of the eccentricity and the inclination  during 
Kozai cycles.

(2) When the mutual inclination between the two orbits is small,  the argument of pericenter 
of a planet secularly circulates in the quadrupolar potential of the binary.
However, \citet{kozai62}  discovered that if the inner planetary orbit is sufficiently inclined
with respect to the binary orbit such that $h < 0.6$, there is a libration solution possible for the planetary  
orbit; the planet's pericenter argument secularly oscillates around either one of the two fixed points, 
$\pm 90^\circ$.  If  the octupole term is included in the perturbing Hamiltonian, 
the pericenter argument in fact is permitted to alternate between circulation and libration  \citep{ford00}. 
The critical Kozai angle corresponding to $h = 0.6$ is $\Ikoz = 39.23^\circ$   
for a planet with  initially small orbital eccentricity. 

(3)  If the initial mutual inclination $I_0$ is above the critical Kozai angle (including a retrograde planetary orbit)
such that $I_0 \in \left[\Ikoz, 180^\circ-\Ikoz \right]$,  then the orbital eccentricity of the planet 
secularly oscillates with a large amplitude.  The evolution of the 
orbital inclination is coupled to the eccentricity oscillation through the Kozai integral,  
and $I$ oscillates between $\Ikoz$ and $I_0$.

(4) The averaged quadrupolar Hamiltonian $\bar{\Hamq}$ is conserved since it is independent of time.
Because the semi-major axis is also constant, a conserved quantity can be defined from the Equation~(\ref{ave_Hamq}),
\begin{equation}
  C \equiv 2+3e_1^2-(3+12e_1^2- 15e_1^2 \cos^2{\omega_1}) \sin^2{I} =  constant.
\end{equation}
This implies  that the maximum orbital eccentricity of the planet $e_{\rm 1, max}$ occurs
when $\omega_1 = \pm 90^\circ$.
Thus,  $e_{\rm 1, max}$ can be written as a function of $C$ and $h$, which are determined from the 
initial condition.  If we assume that the planetary orbit is initially nearly circular so that $e_{1,0} \ll 1$, 
then  $C$ and $h$ are dependent only on the initial mutual inclination $I_0$, and the maximum
orbital eccentricity can be predicted by a simple formula
\begin{equation}
e_{\rm 1, max} \approx \sqrt{1 - \frac{5}{3}\cos^2{I_0}}.
\end{equation}
This approximation is sufficiently accurate in the limit $e_{1,0} \la 0.2$ \citep{fabrycky07}.

\begin{figure}
  \begin{center}
    \plotone{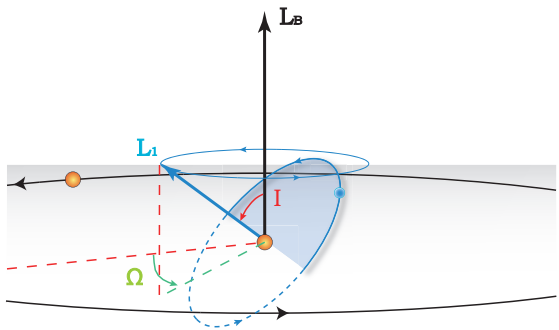}
    \caption{Nodal precession of a planet forced by a binary perturbation.
	As the nodal line of the planetary orbit rotates in the orbital plane
	of the binary, the angular momentum vector of the planet ${\bf L_1}$ 
	sweeps a cone shape around the angular momentum vector of the binary
	${\bf L_{\rm B}}$.  If the initial inclination $I_0$ is above the 
	critical Kozai angle $\Ikoz$, the mutual inclination $I$ also oscillates
	roughly twice between $I_0$ and $\Ikoz$ during one nodal circulation cycle.
	\label{nodalevolution.fig}}
  \end{center}
\end{figure}

(5) The longitude of ascending node of the planet secularly circulates, completing one cycle while 
the orbital inclination and the eccentricity oscillate roughly twice.
This is because of the mirror symmetry of the planetary orbit with respect to the $I =90^\circ$ plane.
The rotation of the planet's nodal line in the binary plane is equivalent to and better visualized
as the precession of the orbital angular momentum vector $\Lin$ around the normal of the reference
plane (= the orbital angular momentum vector of the binary, $\LB$).  Figure~\ref{nodalevolution.fig}
illustrates the nodal precession of a planetary orbit in a binary.  The angle $I$ between the 
two vectors $\Lin$ and $\LB$ corresponds to the mutual inclination between the orbits.  

The nodal precession of a planet in the Kozai solution is always  monotonic circulation 
unlike the pericenter precession, which may alternate between libration and circulation.  
This is because $\Omega_1$ is measured only from an arbitrary reference direction whereas
for $\omega_1$ there exists a fixed line of reference which is the nodal line 
of the planetary orbit in the binary plane.
Thus, the circulation of the planetary ascending node naturally results from secular
three-body interaction and it does not require any critical angle.

(6) The characteristic timescale of the Kozai mechanism can be estimated as \citep{kiseleva98}
\begin{equation}
\tau_{\rm Koz} \approx \frac{2}{3 \pi}\frac{\PB^2}{P_1} \left( 1-\eB^2 \right)^{3/2}  
\frac{m_0 + m_1 + \mB}{\mB}.			\label{tkoz}
\end{equation}
The orbital elements $e_1$, $I_1$, $\omega_1$, and $\Omega_1$ all evolve roughly on this timescale
(except that the precession timescale of the ascending node is $\tau_{\rm koz}$).  
This is why it is often called the ``Kozai resonance'', although it is not strictly a 
resonant interaction \citep{kinoshita99}.

(7)  When there is another source of secular perturbation for the
planetary orbit on a timescale shorter than $\tau_{\rm Koz}$,  this can suppress 
the Kozai cycles.   Competing perturbations include 
general relativistic (GR) precession, tidal and rotational bulges of the 
host star, and other planets in the system \citep{wu03,fabrycky07}.  
If any one of these perturbations causes the pericenter  of a planet to precess
faster than does the Kozai mechanism, then the weak secular torque from the 
binary companion secularly averages to zero, and the Kozai cycles become completely suppressed.


\subsection{Secular Gravitational Coupling Between Planets \label{LLTheory}}

The secular evolution of a pair of planets through mutual gravitational perturbations
has been well-formulated in the second-order Laplace-Lagrange secular 
theory (``L-L theory'', hereafter).  The L-L theory also employs orbital averaging and 
elimination of short-period terms to isolate the secular terms in the disturbing potential.
However, the L-L theory does not assume any hierarchy in the planetary system as does 
the Kozai theory.   The L-L theory is valid for any value of $\alpha \equiv a_1/a_2$ 
as long as the orbital separation between the planets is sufficiently large so that 
there is no short-term perturbation.

In the L-L theory the simplified disturbing function for each planet is derived by 
eliminating the terms that are dependent on the mean longitudes (which vary on short timescales) and 
also the terms that are dependent only on the semi-major axes (which remain constant).
Up to the second order in the eccentricities and the inclinations, the eccentricity and 
inclination perturbations are decoupled and thus the disturbing functions can be written separately
\citep[in the fourth-order terms the evolutions of $e$ and $I$ become coupled when they are 
large; see ][]{veras07}.  It should be also noted that the classical L-L theory becomes 
less accurate in the regime with large orbital eccentricities or inclinations due to the
neglected higher-order terms \citep[higher-order or semi-analytical secular theories have been
recently formulated to improve the accuracy of the classical L-L theory for more general 
planetary orbits; see][]{michtchenko04, libert05}.  
In this paper we employ the L-L theory to derive the approximate evolutionary timescales
of relevant orbital elements of planets.  For this purpose, the classical second-order L-L theory
is still sufficiently accurate, as shown in the comparisons with numerical results in the 
next section.  

The second-order L-L theory has its advantage in that the secular equations of motion of the
planets can be easily derived within a simple eigenvalue problem \citep{brouwer61,murray99}.  
The $e$--$\varpi$ part of the disturbing functions yields two eigenfrequencies, $g_+$ and $g_-$, 
which are functions of the planetary masses and semi-major axes \citep{zhou03}.  
The eigenvalues $g_+$ and $g_-$ are the characteristic frequencies at which the planets' orbital
elements evolve.  For example, in the secular solutions orbital eccentricities of the 
planets oscillate with a period $\tau = 2 \pi /  \left|g_+ - g_-\right|$.
The solution for the longitudes of pericenter is more complex as it is a 
non-linear combination of the two eigenmodes.  However, it can be easily shown 
from the L-L solutions that in the following limit
\begin{equation}
\xi \equiv \frac{\alpha}{1 - 3 q \sqrt{\alpha} / b^{(1)}_{3/2} } \ll 1  \label{LLapprox}
\end{equation}
where $\alpha = a_1/a_2$, $q = m_2/m_1$, and $b^{(1)}_{3/2}$ is the standard Lapalace coefficient 
\citep[Eq.\,7.13 of ][]{murray99},  the $g_-$--mode corresponds to the simple pericenter precession 
of the more massive planet, and $g_+$--mode corresponds to the pericenter precession of the less massive planet.
Because we are mainly interested in  hierarchical planetary systems in which 
the secular perturbation from the stellar companion to $m_2$ dominates over the perturbation from $m_1$ to $m_2$, 
it turns out that $\xi$ is at most 0.1 or typically much smaller for all the systems discussed in 
$\S$~\ref{dynamicaloutcomes}. 
Thus, we will use the derived eigenfrequencies to provide  estimates for the pericenter precession timescales 
$\tppin$ and $\tppout$ for the inner and outer planets, respectively,  as

\begin{eqnarray}
	\left.
	\begin{array}{lll}
 		\tau_{\rm pp, i} & = & 2\pi/g_+ \\ 
		\tau_{\rm pp, j} & = & 2\pi/g_- \\
		\end{array} 	
	\right\}
 	\hspace{0.2cm}{\rm with } \hspace{0.2cm}& m_i < m_j \; (i,j = 1,2; \; i\neq j). \hspace{0.5cm}
\end{eqnarray}

\begin{figure}
  \begin{center}
    \plotone{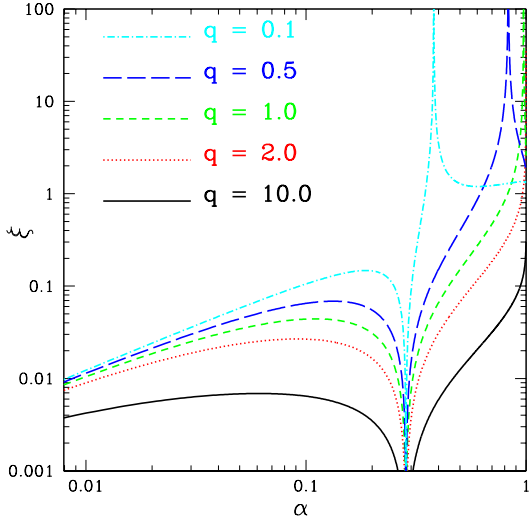}
    \caption{The parameter $\xi$ described in Eq.~(\ref{LLapprox}) as a function of the semi-major
	axis ratio $\alpha$ for different mass ratios $q \equiv m_2/m_1$.  If $\xi \ll 1$, the 
	eigenfrequencies $g_+$ and $g_-$ derived from the L-L theory correspond to the 
	pericenter precession frequencies of the planets.  
	For all the systems discussed in $\S$~\ref{dynamicaloutcomes},
	$\xi \ll 0.1$. \label{eigenfreq.fig}}
  \end{center}
\end{figure}

Likewise, the $I$--$\Omega$ part of the disturbing functions yields two eigenvalues $f_1$ and $f_2$
which describe the characteristic evolutionary frequencies of the orbital inclinations and
the ascending nodes.  Note that there is a subtle difference compared to the solution in the 
$e$--$\varpi$ space;  two planets around a spherical star have only one secular frequency in 
the $I$--$\Omega$ space (e.g., $f_1 = 0$).
This is because the orbital inclination is a relative quantity that requires 
an arbitrary reference plane, whereas the orbital eccentricity is 
a well-defined quantity and introduces asymmetry in the coordinate system.  
There is in fact only one physically meaningful inclination angle for two planets, which is their
mutual inclination.    If, however, the oblateness of the star
is introduced, it would break the degeneracy of the problem, and the L-L theory would yield
two eigenfrequencies in the $I$--$\Omega$ space.  We neglect the quadrupolar perturbation
from the central star and thus derive the single nodal precession timescale $\tppnode$ of 
the planets as 
\begin{eqnarray}
\tppnode = 2\pi / f_2 & & (f_2 \neq 0).  \label{tppnode}
\end{eqnarray}

\subsection{Apsidal and Nodal Coupling of Planets}
A pair of planets around a single star can undergo either  independent pericenter precessions
(apsidal circulation) or coupled pericenter precessions (apsidal libration) \citep{ji03a,barnes06,malhotra02}.
 If the planets follow apsidal libration, then $\Delta \varpi(t) \equiv \left| \varpi_1(t)-\varpi_2(t)
\right|  < C_{\rm amp} + C_{\rm fixed}$  at any given time $t$,  where  $C_{\rm amp} < \pi/2$  is 
the libration amplitude.  There are two different modes of apsidal libration: 
$C_{\rm fixed} = 0$ corresponds to the aligned libration,
and $ C_{\rm fixed} = \pi$ corresponds to the anti-aligned libration.  
Using the secular solutions of the L-L theory, \citet{zhou03} have provided 
the explicit analytical criteria for the occurrence of apsidal librations
given the initial orbital parameters as 
\bea
	\left.\frac{e_2}{e_1}\right|_{\rm initial} & < & 
	-\frac{5}{2} \frac{ \mu \alpha^{3/2} \left(1 - \alpha^2/8\right)  } {1-\mu \alpha^{1/2}} 
	\cos{\Delta\varpi_0} \label{down},
\eea
or, 
\bea
	\left.\frac{e_2}{e_1}\right|_{\rm initial} & > &
	\frac{2}{5} \frac{1- \mu \alpha^{1/2}}{\alpha \left( 1 - \alpha^2/8 \right)} 
	\frac{1}{\cos{\Delta \varpi_0}} > 0,   \label{up}
\eea
where $\mu \equiv m_1/m_2$ and $\Delta \varpi_0 \equiv \left| \varpi_1-\varpi_2\right|_{\rm initial}$.
\citet{zhou03} called the libration regions defined by Equations~(\ref{down}) and (\ref{up})
down- and up-librations, respectively.  

For two planets orbiting around a single star, whether the system
secularly undergoes apsidal libration or circulation can be determined from the initial
orbital configurations.  With the perturbation from a stellar companion, 
however, the above criteria need to be evaluated for each snapshot of time, and 
the planets may alternate between  libration and  circulation.

The role of various external perturbations in affecting the secular apsidal precession mode of
 planets have been previously investigated by several authors \citep[e.g., ][]{malhotra02,chiang02,ford05}.  
\citet{chiang02} have analyzed the secular evolution of the outer two planets c and d
of the $\upsilon$\,Andromedae system and discovered that an adiabatic eccentricity excitation to
the outermost planet d caused by a remnant gaseous disk exterior to the planets would naturally
bring the planets c and d into small-amplitude apsidal libration.  
In this dynamical scenario, there are two effects caused by the outer gas disk to the orbit
of the outer planet: the secular precession of $\varpi_{\rm d}$ (and thus change in $\cos{\Delta\varpi_0}$),
and the adiabatic growth of the eccentricity ratio $e_{\rm d}/e_{\rm c}$.   Since $1 - \mu \alpha^{1/2} > 0$ for $\upsilon\,$And c and d,  
these two effects bring the system into the aligned up-libration as defined in Eq.~(\ref{up}).
Once this up-libration criterion is met, the two planets remain on libration trajectories, maintaining 
$\Delta\varpi < \pi/2$.  As the eccentricity growth of the outer planet continues, the libration 
amplitude further damps down toward $\Delta\varpi = 0$.

Similar apsidal capture mechanism may follow when the outer planet of a double-planet system
is perturbed by a binary companion through the Kozai mechanism.  Because the condition that
the Kozai cycles on the outer planet not be suppressed favors  orbital configurations
such that $1-\mu\alpha^{1/2} > 0$, the most common result from a binary perturbation is the  
 capture into aligned up-libration (Eq.~(\ref{up})).  We present specific examples of 
different apsidal precession modes as well as their effects on the eccentricity evolution of 
planets in binaries in more detail in $\S$~\ref{rigidrotation}.

The ascending nodes of a pair of planets can also secularly circulate or librate. 
The analytical  nodal libration criteria  are  \citep{zhou04}
\begin{equation}
\left. \frac{I_2}{I_1} \right|_{\rm initial} < \frac{2\mu\alpha^{1/2}}{\mu\alpha^{1/2}-1} \cos{\Delta \Omega_0},
\end{equation}
or,
\begin{equation}
\left. \frac{I_2}{I_1} \right|_{\rm initial} > \frac{1}{2} (1-\mu\alpha^{1/2}) \frac{1}{\cos{\Delta \Omega_0}}
\end{equation}
where $\Delta \Omega_0 \equiv \left| \Omega_1 - \Omega_2 \right|_{\rm initial}$.  As previously mentioned,
without the presence of external perturbations, the ascending nodes of two planets precess at one 
characteristic frequency.  If another source of nodal perturbation, for example a secular perturbation 
from a stellar companion, introduces additional nodal precession to one of the planets, then the nodal 
offset $\Delta \Omega$ of the planets may also secularly circulate or librate.

\begin{figure}
  \begin{center}
	\epsscale{0.8}
    \plotone{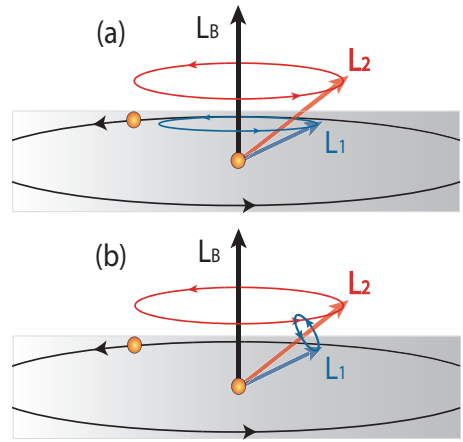}
    \caption{The precession of the orbital angular momentum vectors of double-planet systems in binaries
	that are  (a) in nodal circulation and  (b) in nodal libration.  
	\label{nodalprecessions.fig}}
  \end{center}
\end{figure}

Figure~\ref{nodalprecessions.fig} illustrates the motion of the orbital angular momentum vectors for nodally
circulating and librating systems.  In a system in which the nodal coupling between the planets is weak
compared to an external perturbation, the orbital angular momentum vectors $\Lin$ and $\Lout$ independently 
precess around the normal of the reference plane ($= \LB$ for the case of planets in a binary).  
Notice that in a nodally circulating system, the mutual  inclination angle $I_{12}$  
between the two planetary orbits determined as 
\begin{equation}
\cos{I_{12}} = \cos{I_1}\cos{I_2} + \sin{I_1}\sin{I_2}\cos{(\Omega_1 - \Omega_2)} \label{mutualinclination}
\end{equation}
oscillates within a large range, and the planetary orbits no longer remain coplanar.  
In a nodally librating planetary systems, 
the maximum mutual inclination angle  is limited by the nodal libration amplitude.  In 
Figure~\ref{nodalprecessions.fig}~(b), $\Lout$ secularly circulates around $\LB$ due to the perturbation 
from the binary while $\Lin$ precesses around $\Lout$ maintaining a small angle between the two vectors.  
If the nodal coupling between the planets is sufficiently strong and thus the nodal libration amplitude
remains small, the planetary orbits can remain nearly coplanar as they secularly evolve.  
The rate of nodal precession $\sim 1 / \tppnode$ determined from Eq.~(\ref{tppnode}) 
is an important measure to scale the nodal-coupling
strength of planets.  If an external perturber  precesses the ascending node of one of the planets
at a greater rate, large-amplitude nodal libration or nodal circulation will follow, and the 
planets will open a large mutual inclination angle between their orbits.

\section{Dynamical Classes of Double-planet Systems  in  Binaries }  \label{dynamicaloutcomes}

 We find that there are three distinct dynamical classes of  double-planet systems in binaries: 
(i) completely decoupled systems in which planetary orbits are independently affected by the 
binary perturbation;   (ii) weakly coupled systems in which a large mutual inclination 
angle grows between the planetary orbits due to large-amplitude nodal libration;
(iii) strongly coupled, dynamically rigid systems  in which the inclinations and 
the ascending nodes of the planets evolve in concert.

Each dynamical class is presented with a numerical simulation.  
For all the numerical simulations, we have used the integration package MERCURY~6.2 \citep{chambers99}
with the Bulirsch-Stoer mode to accurately account for possible occurrence of close encounters between
planets.  The general relativistic accelerations are also included  in the integrations to treat the additional 
orbital precession of planets when they are sufficiently close to the primary star.  
The initial planetary orbits are selected to be nearly coplanar ($I_{12} < 5^\circ$) and circular ($e < 0.1$),
and the Hill stability criterion \citep{gladman93} is checked to ensure no immediate planet-planet scattering.  
Each simulation is run for at least five Kozai timescales of the 
outer planet ($\sim 5 P_{\rm B}^2 / P_2$).  

\subsection{Evolution of Decoupled  Planets in Binaries \label{indepkozai}}

\begin{figure}
  \begin{center}
    \plotone{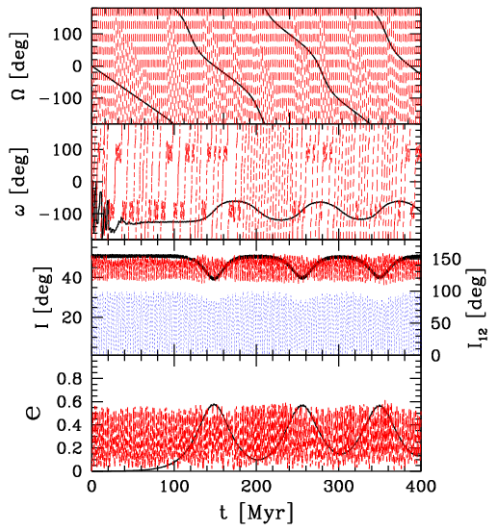}
    \caption{A pair of gravitationally decoupled planets undergoing independent Kozai cycles.  
	The initial orbital parameters of the system are: 
	for the inner planet (slowly varying black curve), $m_1 = 1.0 \MJ$, $a_1 = 2.0\,$AU, 
	$e_1 = 0.01$ and $I_1 = 50^\circ$; for the outer planet (rapidly moving red curve),  
	$m_2 = 0.032 \MJ$, $a_2 = 31.6\,$AU, $e_2 = 0.01$ and $I_2 = 50^\circ$; and for 
	the binary companion,  $\MB = 1.0 \Msol$, 
	$\aB = 750$\,AU and $\eB = 0.20$.  The two ascending nodes are precessing independently.  
	Consequently, the mutual inclination $I_{12}$ (dotted blue curve, scale shown on the right axis) between the planetary 
	orbits oscillates on the Kozai timescale of the outer planet ($\tkozout \sim 2\,$Myr) 
	between $0^\circ$ and $\sim 100^\circ$.  
	\label{indepkozai.fig}}
  \end{center}
\end{figure}

Figure~\ref{indepkozai.fig} illustrates the orbital evolution of a pair of dynamically decoupled
planets in a binary following independent Kozai cycles.  The planetary system is largely hierarchical,
consisting of an inner Jupiter-mass planet at 2\,AU and an outer $\sim 10 \Mearth$ planet at 32\,AU.
Due to the large orbital separation between the planets and the small mass of the outer planet,
the mutual gravitational interaction of the planets is completely suppressed by the perturbation from the 
stellar companion.  
This system can be viewed as two non-interacting test particles placed in a quadrupole potential of a binary, 
resulting in independent Kozai cycles.  Since both planets are initially inclined 
by $50^\circ$ with respect to the binary plane, their orbital 
eccentricities oscillate with roughly equal amplitudes.   The orbital inclinations of the planets 
$I_1$ and $I_2$ measured from  the binary plane also oscillate with the same amplitudes, 
between the initial value of $50^\circ$ and the critical Kozai angle $\Ikoz \sim 40^\circ$.

Notice, however, that the initial coplanarity between the planetary orbits is not maintained at all
throughout the evolution.  Due to the secular perturbation from the stellar companion, 
the orbital angular momentum vector $\Lout$ of the outer planet precesses on a timescale 
$\sim 2 \tkozout \approx 2\,$Myr, while the orbital angular momentum vector $\Lin$ of 
the inner planet precesses similarly but on a much larger timescale $\sim 2\,\tkozin \approx 200\,$Myr  
(see also Figure~\ref{nodalprecessions.fig}~(a)).  
Thus the nodal difference $\Delta\Omega$ between the planets secularly circulates on the timescale
$\tkozout$, causing the mutual inclination $I_{12}$ (which is a function of $\Delta\Omega$, see 
Eq.~\ref{mutualinclination}) to also oscillate on the same timescale with large amplitudes.  
The planetary orbits are maximally inclined with respect to each other every time their
nodes are anti-aligned, reaching $I_{12, \rm max} \approx 100^\circ$, nearly 
twice their initial inclinations with respect to the binary orbit.

In this example the two planets are dynamically invisible to each other since their secular mutual interaction
is  completely suppressed by the perturbation from the stellar companion.
 Thus, even though the mutual inclination $I_{12}$ between their orbits grows beyond 
the critical Kozai angle $\Ikoz$ for a large fraction of time, there is no additional perturbation 
induced to either planet.

\subsection{Weakly Coupled Planets and Induced Eccentricity Oscillations}

Next, we focus on more relevant systems in which (i) the evolution of the outer planet $m_2$ is 
dominated by the secular perturbation from the stellar companion, and (ii) the evolution of the 
inner planet $m_1$ is unaffected by the stellar companion but affected by the secular torque from the
outer planet. 

\begin{figure}
  \begin{center}
    \plotone{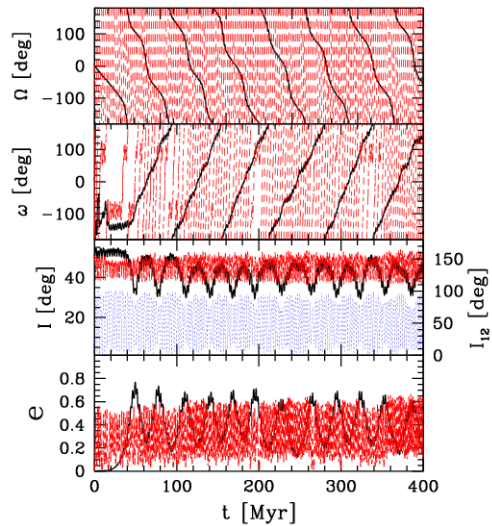}
    \caption{A weakly coupled, nodally circulating double-planet system in a binary.  
	The system is identical to Fig.~\ref{indepkozai.fig} except that the outer planet's mass
	is increased to $m_2 = 0.32 \MJ$.  Due to the increased perturbation by the outer planet,
	the eccentricity of the inner planet oscillates more rapidly compared to Fig.~\ref{indepkozai.fig}.
	\label{acckozai.fig}}
  \end{center}
\end{figure}
The example in Figure~\ref{acckozai.fig} is integrated from the same initial conditions
as those used in Figure~\ref{indepkozai.fig}, except for the outer planet's mass  increased
by a factor of 10 so that $m_2 = 0.32 \MJ$.   The increased perturbation from the outer
planet now suppresses the binary perturbation on the inner planet's orbit.  The nodal precession
periods are $\sim 2\,$Myr for the outer planet (due to the binary) and $\sim 40\,$Myr for
the inner planet (due to the outer planet).  Even though the evolution of the inner planet is
now coupled to the outer planet, the planets are in nodal circulation, i.e., their
orbital angular momentum vectors $\Lin$ and $\Lout$ precess independently, 
and the mutual inclination $I_{12}$ oscillates
between $0^\circ$ and $\sim 100^\circ$, similarly to the previous example.   The difference
here, however, is that the Kozai cycles of the inner planet are introduced by the outer planet 
whose time-averaged inclination with respect to the inner planet is $\sim 50^\circ$.  
Notice that in this case the angular momentum transferred from the binary through the outer
planet accelerates the eccentricity oscillation of the inner planet.  The Kozai timescale
of the inner planet is $\sim P_2^2 / (P_1 m_2) = 40\,$Myr, shorter than 
$\sim 100\,$Myr in the previous example where the inner planet is directly perturbed by the 
stellar companion.  

\begin{figure}
  \begin{center}
    \plotone{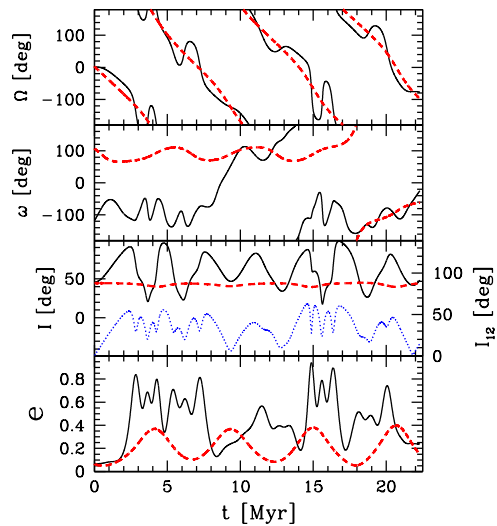}
    \caption{A weakly coupled, nodally librating double-planet system in a binary.  
	The initial orbital parameters of the system are: for the inner planet (solid black curve),  
	$m_1 = 0.02 \MJ$, $a_1 = 0.9\,$AU, $e_1 = 0.07$ and $I_1 = 46^\circ$, for the outer planet 
	(dashed red curve), $m_2 = 0.6 \MJ$, $a_2 = 12.7\,$AU, $e_2 = 0.07$ and $I_2 = 45^\circ$, 
	and for the binary companion,  $\MB = 0.7 \Msol$, $\aB = 803$\,AU and $\eB = 0.78$.  
	The binary  orbit is initially set at $\IB = 0^\circ$.  The dotted blue curve represents 
	the relative inclination angle $I_{12}$  between the two planets (scale shown on the right axis).  
	The nodal offset 
	$\Delta \Omega$ is librating with large amplitudes.  The mutual inclination between the planetary 
	orbits grows chaotically, inducing the large-amplitude eccentricity oscillations to the inner planet.    
	\label{unstable_e.fig}}
  \end{center}
\end{figure}

The secular dynamics of the planets changes drastically 
when the coupling strength between the planets is further increased. 
In Figure~\ref{unstable_e.fig} the outer gas giant  with a mass $m_2 = 0.6 \MJ$ at
$a_2 = 12.7\,$AU follows steady Kozai cycles, unaffected by the inner super-Earth  with 
$m_1 = 6.4 \Mearth$ at $a_1 = 0.9\,$AU.  The binary perturbation to the inner planet is completely suppressed
due to the much stronger secular  torque applied by the outer planet's orbit.   

\begin{figure}
  \begin{center}
    \plotone{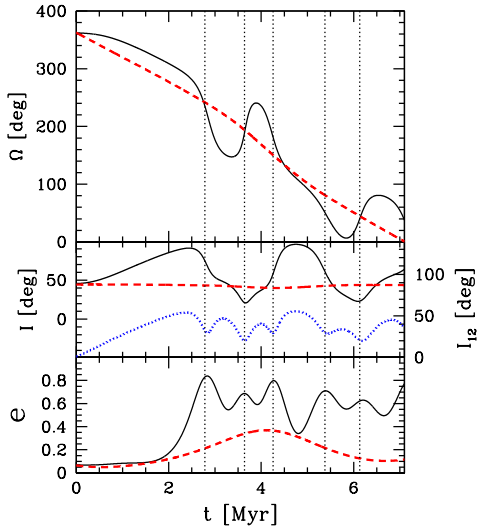}
    \caption{A close-in look of the first 7\,Myr of Figure~\ref{unstable_e.fig}.  
	\label{unstable_e_zoom.fig}}
  \end{center}
\end{figure}

In this system, the two planets secularly follow large-amplitude nodal libration.  
As  illustrated in Figure~\ref{nodalprecessions.fig}\,(b),  in this case 
the trajectory  of the inner planet's orbital angular momentum vector $\Lin$ is a superposition
of two different motions, namely (i) the precession around the binary angular momentum vector $\LB$, 
and (ii) the precession around the outer planet's angular momentum vector $\Lout$.  The resultant motion 
of $\Lin$ is similar to the nutation of the spin axis of a gyroscope.  When the gravitational attraction 
from the outer planet's orbit is not sufficiently strong, $\Lin$ and $\Lout$ can grow a significantly
large mutual inclination angle $I_{12}$ (corresponds to the width of the cone shape that $\Lin$ sweeps around 
$\Lout$), as is the case for the example in Figure~\ref{unstable_e.fig}.  The significant growth
of the mutual inclination  $I_{12}$ induces additional Kozai cycles to the inner planet.  
Each maximum of $e_1$ coincides with the minimum of $I_{12}$, similar to the standard Kozai cycles
(Figure~\ref{unstable_e_zoom.fig}).  However, the amplitude of the induced eccentricity oscillation is not constant 
since the angular momentum of the perturber $m_2$ is also varying due to the interaction 
with the stellar companion.  

Figure~\ref{unstable_e_zoom.fig} focuses on the first 7\,Myr of Figure~\ref{unstable_e.fig}.  
At $t=0$ planets are in nearly coplanar, nodally aligned orbits.  The outer planet's
ascending node starts to rotate in  the  plane of the binary due to the perturbation from
the stellar companion.   As the nodal offset $\Delta \Omega$ between the planetary orbits starts to grow,
due to the gravitational attraction from the outer planet the inner planet begins to 
accelerate its nodal precession.  At the same time, the orbital inclination $I_1$ of the inner
planet begins to grow to conserve the component of its orbital angular momentum parallel to 
$\LB$.  Shortly before 2\,Myr, the mutual inclination $I_{12}$ between the planetary orbits 
reaches $\sim 40^\circ$.  This initiates the Kozai cycles of the inner planet, and $e_1$ starts to 
grow.  At $t \sim2.5\,$Myr the ascending nodes of the planets briefly return to alignment 
and $e_1$ reaches the first maximum.   At this moment, however, the planetary orbits are no longer 
perfectly coplanar because now the inner planet's orbit has been further 
inclined with respect to the binary plane.  The maximally accelerated nodal precession of the inner planet
quickly passes ahead of the nodal line of the outer planet, thereby growing the mutual inclination and 
decreasing the orbital eccentricity.
When the two nodes are maximally separated around $t\sim 3.3\,$Myr, the inner planet's 
nodal precession changes the direction, and $e_1$ starts to grow again.
At $t\sim 3.6\,$Myr the planetary orbits are again nodally aligned,  
completing one nodal libration cycle.  The chaotic eccentricity oscillation of 
the inner planet ensues quasi-periodically,
corresponding to half the nodal libration period of the system estimated from the L-L theory, $\tppnode/2
\approx 2\,$Myr.  
  
We make an additional comment on the effect of the GR precession on the evolution of the inner planet
seen in Figure~\ref{unstable_e.fig}. 
During the sequence $t=8$--14\,Myr, $e_1$ evolves non-periodically, which was not seen in a separate simulation 
of the same system without the GR effect.  Notice that the pericenter argument $\omega_1$ of the inner planet 
departs from the anti-alignment with $\omega_2$ of the outer planet at $t\sim8\,$Myr as the GR precession 
accelerates the precession of $\omega_1$ toward apsidal alignment.  During this epoch of 
the aligned apsidal libration, the induced Kozai cycles on the inner planet are suppressed.  After $t\sim15\,$Myr, 
the system returns to apsidal anti-alignment, and the chaotic oscillations of $e_1$
resumes.

\begin{figure}
  \begin{center}
    \plotone{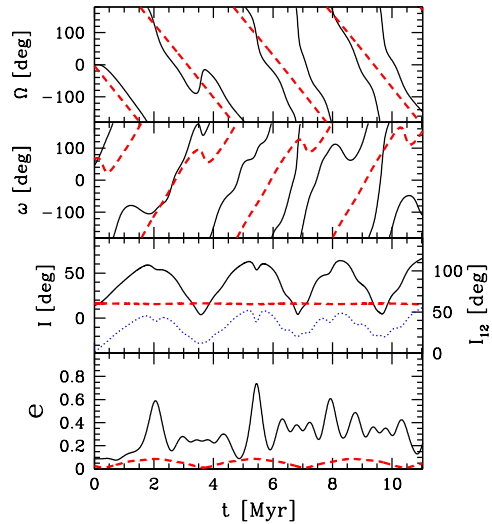}
    \caption{Large-amplitude eccentricity oscillations induced to the inner planet 
	of a double-planet system with initial orbital inclination below the Kozai limit $\Ikoz$.
	The initial orbital parameters of the system are: for the inner planet (solid black curve)  
	$m_1 = 0.02 \MJ$, $a_1 = 0.52\,$AU, $e_1 = 0.09$ and $I_1 = 14.6^\circ$; for the outer planet 
	(dashed red curve)  $m_2 = 0.76 \MJ$, $a_2 = 7.53\,$AU, $e_2 = 0.03$ and $I_2 = 16.5^\circ$; 
	and for the binary companion  $\MB = 0.32 \Msol$, $\aB = 419$\,AU and $\eB = 0.81$.
	Despite the initially small orbital inclination, the eccentricity of the inner planet
	still oscillates with large amplitudes due to the large mutual inclination angle introduced
	between the two planetary orbits, while the Kozai cycles on the outer planet is significantly suppressed.
	 \label{unstable_e2.fig}}
  \end{center}
\end{figure}
The chaotic eccentricity oscillation of the inner planet results from 
the stellar companion introducing a difference in the nodal precession 
frequencies between the planets.  As previously mentioned in $\S~\ref{kozai}$, the nodal 
precession of the outer planet naturally follows from the secular three-body interactions
with the binary stars and thus does not require the critical Kozai angle.  
In the sample system shown in Figure~\ref{unstable_e2.fig}, the two planets are 
initially inclined by only $\sim15^\circ$ with respect to the plane of the binary, significantly 
below  the critical Kozai angle.
Nonetheless, the nodal precession forced on the outer planet's orbit causes the large-amplitude
nodal libration of the planets, leading to the periodic growth of the mutual inclination angle.
The inner planet's orbital eccentricity grows to as large as $e_1\sim 0.8$ while
the eccentricity oscillation of the outer planet is significantly suppressed.

\subsection{Dynamically Rigid Systems \label{rigidrotation}}

The previous examples highlight the evolution of planetary systems in which 
the nodal precession rate of the inner planet $\dot{\Omega}_1$ is small 
compared to the nodal precession rate of the outer planet $\dot{\Omega}_2$ induced by the stellar companion.
Here we turn our attention to  the regime  $\dot{\Omega}_1 > \dot{\Omega}_2$ so that   
the inner planet can effectively couple its 
nodal evolution to that of the outer planet, 
thereby maintaining a small mutual inclination angle.  
\begin{figure}
  \begin{center}
    \plotone{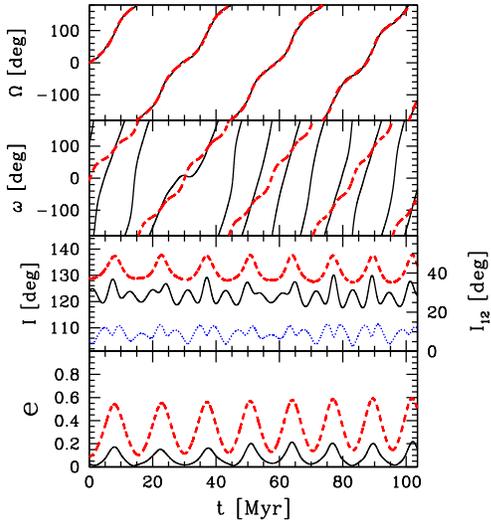}
    \caption{Evolution of a dynamically rigid double-planet system in a binary.  
	The initial orbital parameters of the system are: for the inner planet (black solid curve)
	$m_1 = 0.06 \MJ$, $a_1 = 0.7\,$AU, $e_1 = 0.02$ and $I_1 = 124^\circ$; for the outer planet 
	(red dashed curve)  $m_2 = 0.22 \MJ$, $a_2 = 9.1\,$AU, $e_2 = 0.09$ and $I_2 = 129^\circ$;
	and for the binary companion  $\MB = 0.93 \Msol$, $\aB = 952$\,AU and $\eB = 0.53$.  
	The blue dotted curve shows the relative inclination angle $I_{12}$ between the two planets (scale
	shown on the right axis).  
	In this system  the ascending nodes of the planets are strongly coupled, and the system remains 
	nearly coplanar.
	The pericenter arguments of the planets align briefly near the eccentricity maxima of the outer planet, 
	resulting in the synchronous eccentricity oscillations of the planets. \label{rigidrotation.fig}}
  \end{center}
\end{figure}
Figure~\ref{rigidrotation.fig} illustrates the strong nodal coupling of planets.  
The inner Neptune-size planet ($m_1 = 0.06 \MJ$) orbits
at $a_1 = 0.7\,$AU, while the outer giant planet ($m_2 = 0.22 \MJ$) orbits at $a_2 = 9.1\,$AU.
The nodal precession  timescale of the outer planet is $\sim2 \tkozout = 20\,$Myr.
The secular nodal precession timescale of the inner planet can be estimated from the L-L theory 
and is smaller, $\tppnode \sim 7\,$Myr.  As a result the inner planet successfully follows the nodal precession of the 
outer planet, maintaining tight nodal alignment and thus also a small mutual inclination angle.  
Viewed from the orbital plane of the binary, the  planetary orbits precess in concert, as
if they were embedded in a rigid disk.  We call such systems that rigidly respond to 
external perturbations ``dynamically rigid'' systems.

\begin{figure}
  \begin{center}
    \plotone{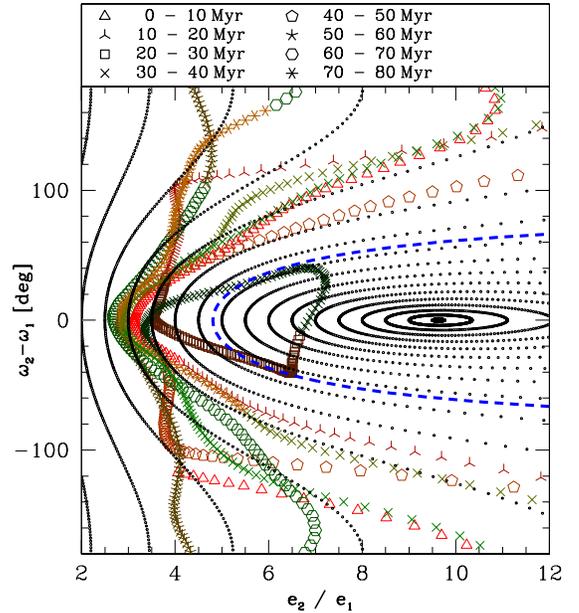}
    \caption{Orbital evolution of the planetary system from Figure~\ref{rigidrotation.fig}, presented in 
	the $e_2/e_1$--$\Delta\omega$ space.   
	The small dots are the secular trajectories, and the blue dashed line is the
	secular separatrix dividing the regions of libration and circulation, both derived from the L-L theory.  
	Different symbols are used to represent the different evolutionary stages of the system computed
	from numerical integration, with each symbol separated by $10^5\,$yr.  The system spends
	a large fraction of time in the region of near apsidal-alignment (small $\Delta\omega$), where
	the eccentricity ratio $e_2/e_1$ of the planets is nearly fixed.  
	During $t\sim20$--$40\,$Myr, the system briefly remains in the apsidal libration regime.}
	\label{contour.fig}  \end{center}
\end{figure}
As seen in Figure~\ref{rigidrotation.fig}, even though the dynamically-rigid planetary orbits  remain nearly coplanar,  
the orbital eccentricity of the inner planet is still  periodically  excited.
In this particular system the inner planet's eccentricity oscillation is coupled to the 
Kozai cycles of the outer planet due to the periodic apsidal capture.  The evolution of the 
same system in the $e_2/e_1$--$\Delta\omega$ space is presented in Figure~\ref{contour.fig}.
The contours with small dots separated with equal time intervals are the secular 
trajectories analytically derived from the L-L theory.
Without the presence of external perturbations,  the planetary system would evolve on 
one of the secular trajectories selected from the initial conditions.  

The planetary system is initially placed in the secular-circulation
region, $(e_2/e_1)_{\rm initial} = 4.5$ and $\Delta\omega_{\rm initial} = -117^\circ$.  As the apsidal offset
$\Delta\omega$ between the planets circulates for $360^\circ$ taking   $\tppin \approx 7\,$Myr, 
the system also shifts rightward in the diagram to new circulation contours due to the eccentricity 
growth of the outer planet (and thus growing $e_2/e_1$) on a timescale $\tkozout \approx 20\,$Myr.  
Note that the contours to the right of the phase space are progressively more concentrated 
near  the apsidal alignment $\Delta\omega \sim 0^\circ$.  
This is because the inner planet's pericenter precesses much faster when its orbital eccentricity is
smaller ($e_2/e_1$ greater).  Since the eccentricity ratio $e_2/e_1$ is roughly fixed to a factor of a few
around the apsidal alignment, at each episode of apsidal alignment the orbital eccentricity of 
the inner planet is excited.  Notice that the eccentricity excitation timescale of the inner planet
is significantly shortened by the outer planet bridging the angular momentum exchange from the 
stellar companion.  Without the presence of the outer planet, the eccentricity excitation from 
the stellar companion occurs on a timescale $\sim 1\,$Gyr and thus would be completely suppressed by the 
GR precession.  With the outer planet aiding the angular momentum exchange between the companion star
and the inner planet, however, the inner planet can evolve on only $\sim 20\,$Myr, 
the timescale of the outer Kozai cycles.

\begin{figure}
  \begin{center}
    \plotone{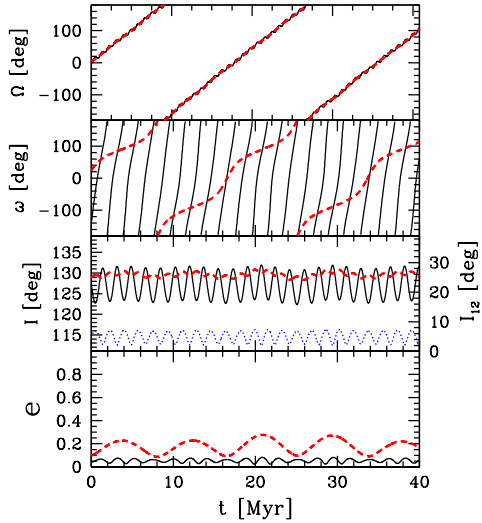}
    \caption{A dynamically-rigid planetary system with decoupled eccentricity evolutions.  
	 The initial orbital parameters of the system are: for the inner planet (black solid curve), 
	 $m_1 = 0.75 \MJ$, $a_1 = 1.15\,$AU, $e_1 = 0.04$ and $I_1 = 127^\circ$; for the outer planet 
	 (red dashed curve),  $m_2 = 1.80 \MJ$, $a_2 = 15.1\,$AU, $e_2 = 0.10$ and $I_2 = 129^\circ$; 
	 for the binary companion, $\MB = 0.80 \Msol$, $\aB = 1322$\,AU and $\eB = 0.72$.  
	The secular torques are applied to the outer planet  by the inner planet and also by the stellar companion
	at similar rates, thereby suppressing the amplitude of the outer planet's eccentricity 
	oscillation.
	The inner planet's pericenter argument precesses  $\sim10$ times as fast as that of the outer planet,
	prohibiting the capture into apsidal alignment.  Consequently, the inner planet's orbital eccentricity
	remains small throughout the evolution. 
	\label{rigidrotation2.fig}}
  \end{center}
\end{figure}

This periodic apsidal-capture mechanism requires the dynamical rigidity of the system 
and a sufficiently large eccentricity oscillation amplitude of the outer planet, as well as comparable apsidal 
precession rates of the planets. 
The sample system shown in Figure~\ref{rigidrotation2.fig} does not satisfy the last two of these conditions
and thus fails to excite the inner planet's eccentricity.  First, in this example 
the two competing secular perturbations from the inner planet and from the stellar companion
both precess the outer planet's orbit in $\sim 20\,$Myr.  As a result, the outer planet's
Kozai cycles are suppressed; its eccentricity reaches only as high as $\sim0.25$ while the 
maximum eccentricity predicted from the initial inclination is 0.58.  Also,  the apsidal precession timescale 
for the inner planet is $\tau\sim 2\,$Myr, much shorter than that of the outer planet, $\tau \sim 20\,$Myr, 
making the episodes of apsidal alignment too brief for the inner planet to adjust its orbital eccentricity.

\begin{figure}
  \begin{center}
    \plotone{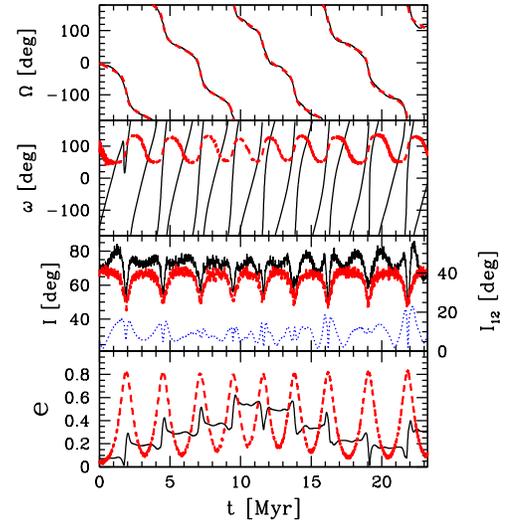}
    \caption{Evolution of a dynamically rigid planetary system in near secular apsidal resonance.
	The initial orbital parameters of the system are: for the inner planet (black solid curve), 
	$m_1 = 1.7 \MJ$, $a_1 = 0.7\,$AU, $e_1 = 0.07$ and $I_1 = 68^\circ$; for the outer planet 
	(red dashed curve),  $m_2 = 4.5 \MJ$, $a_2 = 16.4\,$AU, $e_2 = 0.03$ and $I_2 = 65^\circ$;
	and for the binary companion,  $\MB = 0.4 \Msol$, $\aB = 553$\,AU and $\eB = 0.37$.  
	The pericenter arguments of the two planets evolve on similar timescales. 
	At every Kozai cycle of the outer planet, the pericenter of the inner planet is strongly 
	captured into alignment or anti-alignment with the outer planet, resulting in impulsive excitation
	or damping of $e_1$.
	\label{seculardrift.fig}}
  \end{center}
\end{figure}

In an opposite case in which the two apsidal frequencies are comparable within a factor of a few, 
the orbital eccentricity of the inner planet can be significantly altered 
at each maximum of the outer planet's Kozai cycles.
In Figure~\ref{seculardrift.fig}, $\omega_1$ and $\omega_2$ precess due to the outer planet
and the stellar companion, respectively, both on  timescales of a few Myr. 
These similar apsidal precession frequencies allow the long-lasting tight apsidal alignment 
of the planets during each Kozai cycle of the outer planet.  Analogous to 
the example in Figure~\ref{contour.fig}, as $e_2$ begins to rise in the first 2\,Myr, the 
growing eccentricity ratio $e_2/e_1$ places the system on one of the more elongated circulation contours, 
on which the system quickly returns to the small value of $e_2/e_1$ thereby steeply exciting
$e_1$ at the first Kozai maximum.  When the outer planet's eccentricity begins to decrease, it occurs
on a timescale comparable to the circulation timescale of $\omega_1$, thus the inner planet does
not have enough time to adjust to the circularization of the outer planet.  
Consequently, the eccentricity ratio $e_2/e_1$ steadily drops to small values.  
When the outer planet completes its first  Kozai cycle, the eccentricity ratio is inverted.
For the first few Kozai cycles $e_1$ is more effectively excited than damped, and it
systematically drifts toward larger values.  At the beginning of  the fifth Kozai cycle of 
the outer planet ($t \sim 11\,$Myr), however, $e_2/e_1$ has been significantly reduced so that
the inner planet gets captured into tight apsidal anti-alignment rather than alignment (see
Eq.~\ref{down}).  During the anti-alignment the secular torque from the outer planet transfers
angular momentum to the inner planet's orbit which in turn reduces $e_1$.  Consequently, the
inner planet's eccentricity secularly drifts between 0 and $\sim 0.6$ 
as seen in Figure~\ref{seculardrift.fig}.


\section{Discussion} \label{discussion}
\subsection{Analytical Boundaries of the Dynamical Classes of Planets in Binaries \label{analyticalcriteria}}

\begin{figure*}
\centering
\plotone{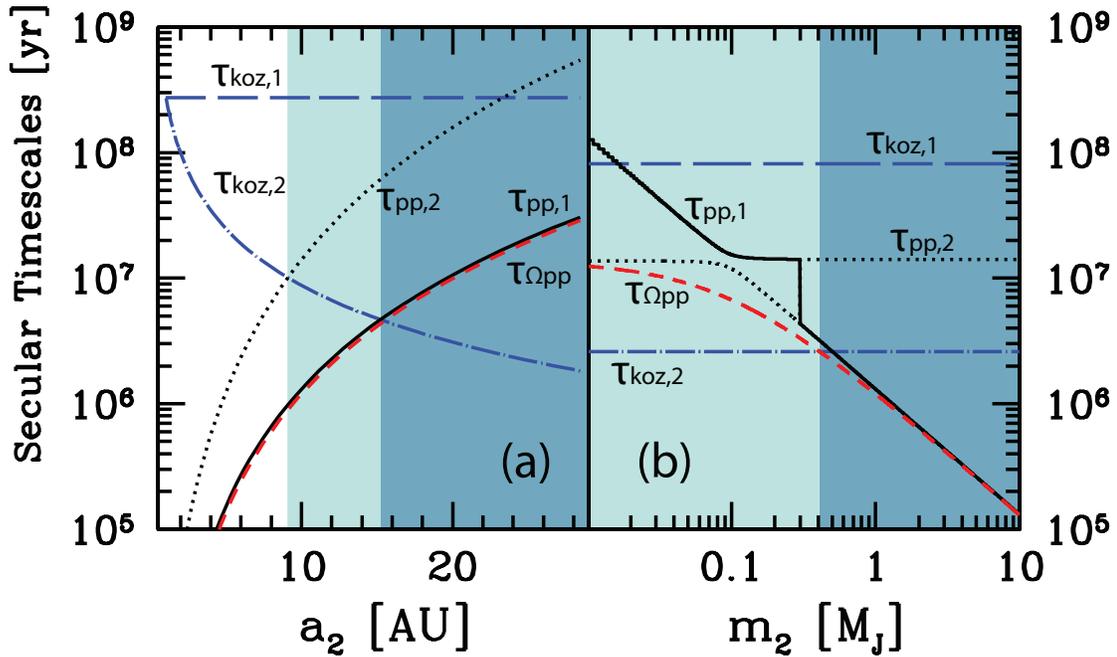}
\caption{Various secular timescales of double planets in a binary as a function of (a) the orbital semi-major axis
and (b) the mass of the outer planet, with the following system parameters : $m_1 = 0.3\MJ$, 
$a_1 = 1\,$AU, $\mB = 1 \Msol$, $\aB = 500\,$AU, and $\eB = 0.5$.  In the left panel (a) $m_2 = 1\MJ$, and
in the right panel (b) $a_2 = 10\,$AU.  
In the unshaded area, the Kozai cycles on the outer planet would be suppressed by 
the secular torque from the inner planet.   In the light-shaded region, planetary systems 
undergo dynamically-rigid evolution.  In the dark-shaded region, the orbital eccentricity of the 
inner planet grows chaotically due to the growing mutual inclination between the planetary orbits.
\label{secular_timescales.fig}}
\end{figure*}


\begin{deluxetable*}{llccccc}
\tabletypesize{\scriptsize}

\tablecaption{Dynamical Classes of Double Planets in Binaries \label{tbl-1}}
\tablewidth{0pt}
\tablehead{
\colhead{Dynamical Class} & \colhead{Description} & \colhead{$\Delta \Omega$} & \colhead{$\Delta I$} 
& \colhead{$\Delta_\omega$} & \colhead{$e_1$} & \colhead{Figure} 
}
\startdata
Independent Kozai Cycles & Planets gravitationally decoupled & Circ. & Oscillates between 0 and $2 I_0$ & Circ. & Kozai cycles  & 
\ref{indepkozai.fig} \\
Nodally decoupled & Accelerated Kozai cycles & Circ. & Oscillates between 0 and $2 I_0$ & Circ. & Kozai cycles 
\tablenotemark{a} & \ref{acckozai.fig} \\

Large-amplitude nodal libration   & Additional chaotic Kozai cycles \tablenotemark{b} & Lib.  & large & 
Lib. \tablenotemark{c}  & Kozai cycles \tablenotemark{d} & \ref{unstable_e.fig}, \ref{unstable_e2.fig} \\

Dynamically rigid & Synchronized Kozai cycles & Lib. & small & Circ. \tablenotemark{e}  & Kozai cycles  & 
\ref{rigidrotation.fig}\\

Dynamically rigid & Unaffected $e_1$ & Lib. & small & Circ.  & small & \ref{rigidrotation2.fig}  \\

Dynamically rigid & Secular $e_1$ drift & Lib. & small & Circ. \tablenotemark{e}  & large & \ref{seculardrift.fig}  \\

\enddata

\tablecomments{Circ. and Lib. indicate circulation and libration, respectively.}
\tablenotetext{a}{caused by the outer planet, provided $I_0 \ga \Ikoz$}
\tablenotetext{b}{also occurs for $I_0 < \Ikoz$}
\tablenotetext{c}{anti-aligned libration for a large fraction of time}
\tablenotetext{d}{with varying amplitude, caused by the outer planet}
\tablenotetext{e}{with brief periodic episodes of apsidal capture}
\end{deluxetable*}


Table~\ref{tbl-1} summarizes the coupling properties of the planets' orbital elements in each dynamical class.
The analytical boundaries in the parameter space of double-planet systems in binaries distinguishing
each  dynamical class presented in the previous section can be provided by comparing the several
competing secular perturbation timescales of the systems.    
The L-L theory yields three eigenfrequencies that can be translated into orbital  evolution timescales  
$\tppin$, $\tppout$, and $\tppnode$.
For a pair of planets with a sufficiently large mass or semi-major axis ratio such that the
outer Kozai cycles are not suppressed,  $\tppin$ and $\tppout$ are approximated to the apsidal precession 
timescales of the inner and the outer planets, respectively, and $\tppnode$ corresponds to the   
characteristic nodal precession timescale of the planets.  A stellar companion also secularly perturbs
the inner and the outer planet on timescales  $\tkozin$ and $\tkozout$, respectively.
Figure~\ref{secular_timescales.fig} shows that, generally, for a large region of the parameter 
space $\tppin < \tkozin$, a regime in which the inner planet's evolution 
is affected only by the outer planet.  
For the outer planet, the condition we have imposed is such 
that the  perturbation from the stellar companion overcomes the perturbation from the 
inner planet, i.e., $\tppout > \tkozout$, typically corresponding to  relatively 
hierarchical systems (the shaded regions in Figure~\ref{secular_timescales.fig}).  

For such hierarchical systems in which the stellar companion secularly perturbs the outer planet's
orbit, two distinctively different evolutionary processes are possible, depending on whether 
the planets can rigidly respond to the quadrupole perturbations from the stellar companion,
maintaining  a small mutual inclination angle $I_{12}$.   The key quantity which separates 
the two dynamical classes is the nodal precession rate of each planet.  The ascending node of the 
outer planet secularly precesses due to the binary companion roughly on a timescale $2\,\tkozout$.  
The nodal-coupling strength
of the inner planet to the outer planet can be quantified by the nodal precession timescale $\tppnode$ 
derived from the L-L theory.  If the outer planet's nodal precession occurs faster than 
that of the inner planet such that
\begin{equation}
	 \tppnode > 2\tkozout\,,
\end{equation}
corresponding to the darker-shaded region of Figure~\ref{secular_timescales.fig}, then a large
mutual inclination angle opens between the planetary orbits, and the additional Kozai cycles
are induced to the  inner planet on a timescale $\tau_{\rm ind, Koz}$ which is shorter 
than the timescale $\tkozin$ of a direct perturbation from the stellar companion, 
\begin{equation}
	\tau_{\rm ind, Koz} \sim \frac{2}{3} \frac{P_2^2}{P_1} \frac{m_0}{m_2} < \tkozout < \tkozin.
\end{equation}
The amplitude of the induced eccentricity oscillation is not constant because 
the Kozai integral of the inner planet measured in the frame of the outer planet's orbit
does not remain constant as the outer planet is also subject to angular momentum exchange 
with the stellar companion.

If, on the other hand, the gravitational nodal coupling between the planetary orbits 
is comparably strong such that the nodal precession of the inner planet does not 
lag significantly behind that of the outer planet, i.e.,
\begin{equation} 
	 \tppnode < 2 \tkozout ,
\end{equation}
then the planetary orbits can remain close to coplanar.  

In dynamically rigid planetary systems the orbital eccentricity of the inner planet may still evolve, 
depending on the strength of the apsidal coupling between the planets.  
The most extreme case is when the two apsidal frequencies
are close to the secular resonance such that
\begin{equation}
	\tppin \sim \tkozout.
\end{equation}
In such cases, the inner planet gets periodically trapped into apsidal alignment or anti-alignment
with the outer planet.  Such repeated apsidal coupling is responsible for the secularly-drifting orbital
eccentricity of the inner planet  illustrated in Figure~\ref{seculardrift.fig}.  
When the pericenter evolution of the planets is less strongly coupled,  
the orbital eccentricity of the inner planet is periodically excited synchronously with 
the outer Kozai cycles (Figure~\ref{rigidrotation.fig}).  If the coupling further weakens such that 
\begin{equation}
	\tppin \ll \tkozout,
\end{equation}
then the two apsidal precessions are completely independent, and the inner planet's eccentricity
remains small (Figure~\ref{rigidrotation2.fig}).  

\subsection{Stability of  Planets in Binaries}
A stellar companion may induce instability to the double-planet system around the primary star
that can otherwise remain stable for a long term.  
A possible trigger of instability is either the Kozai cycles of the outer planet,
or the chaotic eccentricity oscillation induced to the inner planet.
In both cases what may follow is the  loss of a planet either through collision with the host star 
or ejection from the system.
The first case of instability is rarely observed in our simulations because we have imposed  
the Hill stability criterion on the initial conditions, including the predicted maximum orbital
eccentricity of the outer planet.  The majority of the unstable outcomes are found 
in systems with additional Kozai cycles induced to the inner planet due to the broken coplanarity 
between the planetary orbits.  It needs to be remembered, however, that our simulations do not
include any orbital damping mechanism such as tidal dissipation effects.  Because the 
eccentricity growth of the inner planet induced by the outer planet is rather adiabatic
than impulsive,  were tidal effects included in the simulations, many of these highly 
eccentric planets would effectively damp their orbital eccentricities and migrate to tighter  orbits 
before being lost from the system.  

\section{Summary}

We have shown that the complex interplay between the secular gravitational perturbations from a 
stellar companion and the gravitational coupling between planets produces various classes of dynamical outcomes.
The two distinct evolutionary processes are the dynamically-rigid evolution of planetary orbits, and 
the chaotic growth of the mutual inclination angle between the planets.
The dynamical rigidity of planetary systems can be analytically estimated by comparing  
the nodal precession rates of the planets.  A weak nodal coupling between the planets such that $\tppnode
> \tkozout$ leads to large-amplitude nodal libration which  steeply inclines the orbital plane of 
the inner planet with respect to the outer planet.  
Consequently, the inner planet suffers secular eccentricity oscillation 
with chaotically varying amplitudes.  If, on the other hand, the gravitational nodal coupling 
of the planets is stronger than the  nodal perturbation to the outer planet caused by the 
stellar companion such that $\tppnode < \tkozout$, then the orbital planes of the planets 
precess in concert, maintaining near-coplanarity.  Even in dynamically-rigid planetary systems, the 
orbital eccentricity of the inner planet can remain small, or periodically excited in response 
to the outer Kozai cycles, or  secularly drift within a large range,
depending on the  apsidal precession frequencies of the planets.  

 The periodic  eccentricity and inclination perturbations from 
 a stellar companion propagating inward through the planetary system
effectively reduces the orbital evolution timescale of the inner planet.
Such a cascading angular momentum transfer mechanism needs to be incorporated to 
the existing theoretical models of extrasolar planetary systems in binaries.  
The traditionally adopted constraint of the critical Kozai angle between the planetary systems
and the binary orbits may be even relaxed, if undetected, non-coplanar planetary companions are applying  
eccentricity perturbations to the detected ones.  Figure~\ref{secular_timescales.fig} indicates
that there is a range of the orbital radius of the outer planet within which the 
additional Kozai cycles and significant eccentricity excitation of the inner planet 
are almost certain ($a_2 \ga 15\,$AU for the system presented in Figure~\ref{secular_timescales.fig}).
Also, in a less hierarchical system in which dynamical rigidity is ensured
($a_2 \sim 10$--15\,AU in Figure~\ref{secular_timescales.fig}),
the inner planet's eccentricity can be still excited given suitable apsidal-capture conditions.
Thus, the eccentricity excitation of extrasolar planets by binary perturbations is possibly much more efficient than
 previously predicted in the studies based on single-planet assumptions \citep{takeda05}.

In this paper we have provided a general dynamical framework for understanding planetary systems in binaries.
Yet, there remain some  limitations in our study which will be further investigated in our future papers.   
First, a broader range  of initial planetary eccentricities and inclinations
needs to be considered.  Recent theoretical studies have shown that newly-formed planets
emerging from a protoplanetary disk may already possess large orbital eccentricities or
inclinations because of  interactions with the gaseous disk \citep{goldreich03,moorhead07} or
through planet--planet scattering events \citep{chatterjee07,juric07,nagasawa08}.  More realistic distributions
of  initial orbital parameters are necessary to accurately determine the likelihood of each dynamical class
of planets in binaries presented here.  Second, we have not included  tidal dissipation,
which may also affect the dynamical outcomes of planets in binaries.
The orbital energy dissipation of highly eccentric planets through tidal friction and
subsequent orbital migration have been previously studied for hierarchical
three-body systems \citep{wu03,fabrycky07,wu07}. 
Since  significant eccentricity growth of the inner planet naturally results when
the coplanarity between planetary orbits is destroyed, the formation of close-in planets among
multiple-planet systems in binaries through tidal dissipation may be fairly common.

As the observational techniques continue to advance, direct measurements of the
following observable signatures will be possible to constrain the dynamical outcomes 
of planets in binaries:  (1) the mutual inclination angle among multiple-planet systems
in binaries measured by astrometric observations \citep[e.g., the preliminary results for 
$\upsilon$\,Andromedae system by HST Fine Guiding Sensor show a large mutual 
inclination angle $\sim 35^\circ$ between the planets c and d; ][]{mcarthur07};  (2)
large spin-orbit misalignments of close-in transiting planets possibly formed via induced Kozai migration,
measured through the Rossiter-McLaughlin effect \citep{fabrycky07}; (3) detections of additional 
planets beyond a few AU from the primary stars of binary systems by continuous radial-velocity 
monitoring of known planetary systems; (4) detections of stellar companions around known hierarchical 
multiple-planet systems by ongoing co-proper motion surveys.

\acknowledgements

This work was supported by NSF Grant AST-0507727. We are grateful to Lamya Saleh for providing us an
improved version of the Mercury BS integrator including the lowest-order post-Newtonian correction.  
We also thank Ji-Lin Zhou, Daniel Fabrycky, and Fred Adams for insightful discussions.



\end{document}